\documentclass[usenatbib]{mnras} 
\usepackage{graphicx}
\def\lapp{\ifmmode\stackrel{<}{_{\sim}}\else$\stackrel{<}{_{\sim}}$\fi}

\begin{document}

%%paper title
%%For line breaks \\ can be used within title 
\title{Millisecond Pulsars, their Evolution and Applications}

%%author names are separated by comma (,) 
%%use \and before the last author name 
%%use a * along with the number separated by comma
%% for the  author for correspondence
%%\textsuperscript{number} is used for affiliation
%%\affilOne, \affilTwo etc., upto \affilTwentyfive is possible
%%Please note the first letter after \affil is capitalised in the command
%%

\author{R. N. Manchester\\
  CSIRO Astronomy and Space Science, PO Box 76, Epping NSW 1710, Australia\\
  dick.manchester@csiro.au
}

%%escape two column mode for title, affiliation and abstract
%%by giving \twocolumn command as shown

%\twocolumn[{

\maketitle

%%include \corres to print the corresponding author Email id
%\corres{dick.manchester@csiro.au}

%%include \msinfo for
%%manuscript information such as
%%received, revised and accepted dates
%%
%\msinfo{6 March 2017}{26 July 2017}{ 2017}

%%abstract
\begin{abstract}
  Millisecond pulsars (MSPs) are short-period pulsars that are
  distinguished from ``normal'' pulsars, not only by their short
  period, but also by their very small spin-down rates and high
  probability of being in a binary system. These properties are
  consistent with MSPs having a different evolutionary history to
  normal pulsars, viz., neutron-star formation in an evolving binary
  system and spin-up due to accretion from the binary companion. Their
  very stable periods make MSPs nearly ideal probes of a wide variety
  of astrophysical phenomena. For example, they have been used to
  detect planets around pulsars, to test the accuracy of gravitational
  theories, to set limits on the low-frequency
  gravitational-wave background in the Universe, and to establish
  pulsar-based timescales that rival the best atomic-clock timescales in
  long-term stability. MSPs also provide a window into stellar and
  binary evolution, often suggesting exotic pathways to the observed
  systems. The X-ray accretion-powered MSPs, and especially those that
  transition between an accreting X-ray MSP and a
  non-accreting radio MSP, give important insight into the physics of
  accretion on to highly magnetised neutron stars. 
\end{abstract}

%%insert keywords separated by 3 hyphens using \keywords{words}
%}]
%%close the twocolumn escape here

%%include \doinum{number}for the DOI number in the header
%%include \volnum{number} for the volume number in the header
%%include \year{yyyy} for  year of publication in the header
%%include \pgrange{num--num} page range of article in the header
%%include \artcitid{num} for the article citation id
%%include \lp to print last page of the article
%%include \setcounter{page}{pagenum} for the exact starting page of the article

%\doinum{12.3456/s78910-011-012-3}
%\artcitid{\#\#\#\#}
%\volnum{123}
%\year{2016}
%\pgrange{23--25}
%\setcounter{page}{23}
%\lp{25}

\section{Introduction}\label{sec:intro}
The first pulsars discovered \citep{hbp+68,phbc68} had pulse periods
between 0.25~s and 1.3~s. Up until 1982, most of the 300 or so known
pulsars had similar periods, with the notable exceptions of the Crab
pulsar \citep{sr68}, the Vela pulsar \citep{lvm68}, and the
Hulse-Taylor binary pulsar \citep{ht75a}. These had periods of 33~ms,
89~ms and 59~ms respectively. The Crab and Vela pulsars had rapid
slow-down rates showing that were very young and almost certainly
associated with their respective supernova remnants. Discovery and
timing of these and other pulsars led to the conclusion that pulsars are
rotating neutron stars, born in supernova explosions with periods 10
-- 20~ms and gradually slowing down to periods of order 1~s over
millions of years \citep{gol68,rc69b,rckm69,hun69,mp72}.

This cosy consensus was somewhat shaken in 1982 by \citet{bkh+82}
announcing the discovery of the first ``millisecond
pulsar'' (MSP), PSR B1937+21, with the amazingly short period of just
1.558~ms. This pulsar was found in September 1982 at Arecibo
Observatory in a very high time-resolution search of the enigmatic
steep-spectrum compact source 4C21.53W. This source also showed strong
interstellar scintillation and strong linear polarisation. All of
these properties suggested an underlying pulsar, but previous searches with lower
time resolution (including one by the author) had failed to reveal any
periodicity.
  
Within days of the announcement of the discovery of PSR B1937+21,
\citet{rs82} and \citet{acrs82}
proposed that the MSP resulted from the ``recycling'' of an old,
slowly rotating and probably dead neutron star through accretion from
a low-mass companion. The mass transferred from the companion also
carries angular momentum from the orbit to the neutron star, spinning
it up and reactivating the pulsar emission process. This idea built on
earlier work by \citet{sb76} and \citet{sv82} in which the relatively short period 
and large age of the original binary pulsar, PSR B1913+16,
\citep{ht75a,thf+76} and maybe some other binary pulsars \citep{btd82},
were explained by invoking such an accretion process.\footnote{For the
  purposes of this article, we define an MSP to be a pulsar with
  period less than 100~ms and period derivative less than
  $10^{-17}$. The somewhat generous period limit allows recycled
  pulsars such as PSR B1913+16, to be included and the
  period-derivative limit excludes young pulsars such as the Crab and
  Vela pulsars.}
 
The next two MSPs to be discovered, PSR B1953+29 \citep{bbf83} and PSR
B1855+09 \citep{srs+86}, were both members of a binary system,
consistent with the recycling idea. At the time, only five of the more
than 400 ``normal'' (non-millisecond) pulsars known were binary,
compared to three of the four known MSPs. At first glance, the absence
of a binary companion for PSR B1937+21 was surprising, but it was
quickly recognised, even in the Backer et al. discovery paper and by
\citet{rs82}, that this could be explained
by disruption of the binary by asymmetyric mass loss in an
accretion-induced collapse of the likely white-dwarf remnant of the
companion star. With the later discovery of the ``black widow''
pulsar, PSR B1957+20 \citep{fst88}, it was realised that complete
ablation of the companion star was another viable mechanism for
formation of solitary MSPs. 

The first MSP in a globlular cluster, PSR B1821$-$24A in M28, was
discovered in 1987 by \citet{lbm+87}. This set off an
avalanche of discoveries of MSPs in globular clusters, with 21 MSPs
being discovered in globular clusters by 1993, with eight in both M15
\citep{agk+90,and92} and 47 Tucanae \citep{mld+90,mlr+91}. There are now
145 globular-cluster pulsars known, all but a handful of them
MSPs. Clearly globular clusters are efficient factories for the
production of MSPs, see \citep{rpr00,vf14}.

Another important development in MSP research was the discovery that
MSPs are relatively strong emitters of pulsed $\gamma$-rays. Although
predicted by Srinivasan in 1990 \citep{sri90,bs91}, the first
observational evidence was the tentative detection by \citet{khv+00}
of pulsed $\gamma$-ray emission from PSR J0218+4232,
a known binary radio MSP with a period of 2.3~ms, in {\it EGRET} data,
later confirmed as one of eight MSPs detected with the {\it Fermi
  Gamma-ray Space Telescope} \citep{aaa+09f}. About 25 previously known
radio MSPs have now been detected as $\gamma$-ray pulsars by folding
the {\it Fermi} data at the known radio period \citep{egc+13}.

It was soon recognised that $\gamma$-ray detected pulsars had rather
unusual $\gamma$-ray properties compared to other classes of
$\gamma$-ray sources, for example, they are steady emitters over long
intervals and have characteristic power-law spectra with an
exponential cutoff at a few GeV \citep{aaa+13}. Radio searches of
previously unidentified {\it Fermi} sources with these properties have
been extraordinarily successful in uncovering MSPs, with about 50 so
far identified, e.g., \citep{ckr+15,cck+16}. In about half of these,
$\gamma$-ray pulsations have subsequently been detected by folding the
$\gamma$-ray data with the precise period ephemeris from the radio
observations. One particularly interesting aspect of these MSP
discoveries is that many are in short-period binary orbits ($P_b \lapp
1$~day) with low-mass companions (M$_c \lapp 0.3$~M$_\odot$) and
exhibit radio eclipses due to gas ablated from the companion, forming
black widow or redback systems\footnote{The names ``black widow'' and
  ``redback'' were coined by 
  \protect\citet{el88} and  \protect\citet{rob13},
  respectivly, after the rather ungracious female spiders that have a
  tendency to consume their much smaller male companion after
  mating. The pulsar analogy is that, in these close binary systems,
  ablation of the companion star by the pulsar wind may destroy it,
  with no thanks for the fact that earlier accretion from the
  companion star gave the pulsar its rapid spin and energetic wind.}

In parallel with these developments, the wide-field radio searches for pulsars
continued, discovering many MSPs. Particularly successful were the
Parkes Swinburne mid-latitude survey (14 MSPs, see \citep{eb01b}), the
Parkes Multibeam Survey (28 MSPs, \citet{fsk+04}), the Parkes
``HTRU'' surveys (28 MSPs, \citet{btb+15}), the Arecibo
``PALFA'' survey (21 MSPs, \citet{skl+15}) and the Green Bank
low-frequency surveys (15 MSPs, \citet{blr+13,slr+14}). Analysis
or re-analysis of many of these surveys is continuing and more
discoveries can be expected.

\begin{figure}
\includegraphics[width=.95\columnwidth]{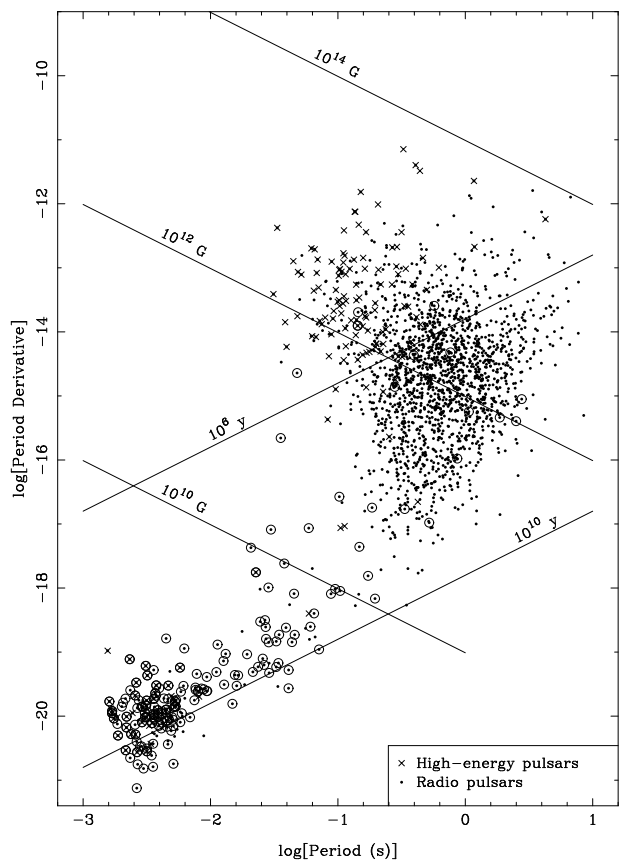}
\caption{Plot of period slow-down rate ($\dot P$) versus pulsar period
  ($P$) for Galactic disk pulsars. Binary pulsars are indicated by a circle around
  the point and pulsars that emit at high energies (optical and above)
  are marked. Lines of constant characteristic age, $\tau_c =
  P/(2\dot P)$, and surface-dipole magnetic field strength, $B_s =
  3.2\times 10^{19}\,(P \dot P)^{1/2}$~G, based on slow-down due to
  magnetic-dipole emission, are also shown. Globular-cluster pulsars
  are not included as the observed $\dot P$ is often significantly
  affected by acceleration in the cluster gravitational field. }\label{fg:ppdot}
\end{figure}

These various searches have revealed a total of 255 MSPs, roughly 10\%
of the known pulsar population. Of these, more than 180 are members of
binary systems, with orbital periods ranging from 1.5 hours (PSR
J1311$-$3430) to nearly 700 days (PSR J0407+1607)\footnote{Other
  longer-period binary systems are known, but in these cases the
  pulsar is probably not recycled.}. By comparison, only 24 or about 1\%
of the normal pulsar population, are binary. Figure~\ref{fg:ppdot}
illustrates the distinct properties of MSPs compared to normal
pulsars, viz., much shorter period, very small slow-down rate and
predominance of binary membership. If one assumes period slow-down due
to emission of magnetic-dipole radiation (electromagnetic waves with a
frequency equal to the pulsar spin frequency) or acceleration of
pulsar winds in a dipole magnetic field, the characteristic age
$\tau_c$ and surface-dipole field strength $B_s$ can be
estimated. $\tau_c$ is a reasonable estimator of the true age of
normal pulsars, but only an upper limit on the true age of MSPs since
it assumes that the pulsar was born with infinite spin frequency with
regular magnetic-dipole slow-down after that. MSPs have a much more
complicated spin history, see \citet{bv91}.

One of the main reasons that MSPs are so important is that their spin
periods are extraordinarily stable. This enables their use as
``celestial clocks'' in a wide variety of applications. Studies of
pulsar ``timing noise'', e.g., \citep{sc10}, show that MSP periods are
typically more than three orders of magnitude more stable than those
of normal pulsars, and for the best cases, e.g., PSR J1909$-$3744, see,
e.g., \citet{hcm+12}, have a stability rivalling that of the best
atomic clocks. This great period stability may be related to
the very weak external magnetic fields of MSPs
(Figure~\ref{fg:ppdot}).

Despite having been proposed as early as 1969 \citep{og69}, the issue
of magnetic field decay in pulsars is not yet resolved, with recent
population studies of normal pulsars, e.g., 
\citet{fk06} and the discovery of low-field but young pulsars, e.g.,
PSR J1852+0040 which is associated with the supernova remnant
Kesteven 79 but has a dipole surface field of only $3\times 10^{10}$~G
\citep{hg10}, suggesting that decay of normal pulsar magnetic fields is
not required. If this remains true over the $\sim 10^9$~yr timescale
for recycling, then the low fields of MSPs must be a by-product of the
recycling process, for example, through burial of the field by
accreted material \citep{pm07}. On the other hand, if fields do decay
on timescales of $10^9$~yr or less, then the problem becomes
accounting for the low but finite field strengths of MSPs. An
innovative solution to this problem in which field decay is related to
the variable spin-down rate across the whole history of the
present-day MSP was presented by \citet{sbmt90}.

In \S\ref{sec:binary} we discuss the use of MSPs as probes of binary
motion, including the detection of planetary companions and the
investigations of relativistic perturbations leading to tests of
gravitational theories. The search for nanoHertz gravitational waves
using pulsar timing arrays (PTAs) is described in \S\ref{sec:pta} and
the use of PTA data sets to establish a pulsar-based timescale is
dicussed in \S\ref{sec:timescale}. The different classes of binary and
millisecond pulsars and their formation from X-ray binary systems
through the recycling process are discussed in \S\ref{sec:evol}. In
the concluding section (\S\ref{sec:concl}) we highlight the rich
research fields opened up by the discovery of binary and millisecond
pulsars and the important contributions of Srinivasan to many aspects
of this work.

\section{MSPs as probes of binary motion}\label{sec:binary}
\subsection{Planets around pulsars}\label{sec:planets}
\begin{figure}
\includegraphics[width=.95\columnwidth]{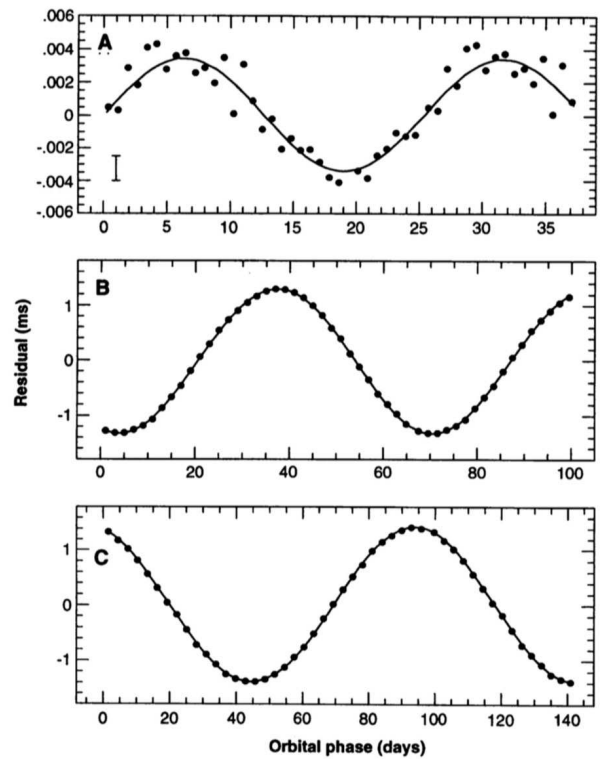}
\caption{Timing signatures of the three planets that orbit the MSP PSR
  B1257+12. The orbital periods are 25.34 days, 66.54 days and 98.22
  days and the planetary masses are about $0.015 M_\oplus$, $3.4
  M_\oplus$ and $2.8 M_\oplus$, for planets A, B and C,
  respectively. \citep{wol94}}\label{fg:B1257}
\end{figure}

The first detection of a planet around a star other than the Sun was
made by \citet{wf92} who discovered two planets
orbiting PSR B1257+12, an MSP with a pulse period of 6.2~ms. The
planets have orbital periods of about 66 and 98 days, are in circular
orbits of radii 0.36 and 0.47 AU and have masses of $3.4/\sin i$ and
$2.8/\sin i$ Earth masses, respectively, where $i$ is the (unknown)
orbital inclination. In 1994,  \citet{wol94} announced
the discovery of a third planet in the system with a mass close to
that of the Moon and an orbital period of approximately 25 days. This
remains (by a wide margin) the least massive planet known for any
star, and its detection is an excellent demonstration of the power of
pulsar timing. Figure~\ref{fg:B1257} shows the timing signatures of
the three planets. The 1994 paper also announced the detection
of predicted small perturbations in the orbital periods of the two
larger planets. This observation unequivocally confirmed that the
timing modulations observed in this pulsar are caused by orbiting
planetary bodies.

PSR B1620$-$26, located in the globular cluster M4, has a pulse period
of about 11~ms and a binary companion of mass about 0.3~M$_\odot$ in a
191-day, almost circular, orbit \citep{lbb+88}. Continued timing
observations showed evidence for additional
perturbations to the pulsar period that could result from the presence
of a third body, possibly of planetary mass \citep{tat93}. Analysis of
an 11-year timing dataset by \citet{tacl99} showed that the results
were consistent with a Jupiter-mass planet with an orbital period of the
order of 100 years. \citet{srh+03} used Hubble Space Telescope
observations to determine a mass for the wide-dwarf companion
which in turn fixed the orbital inclination and constrained the outer
planetary companion to have an orbital radius of about 23 AU and mass
about 2.5 Jupiter masses. 

The third pulsar known to have a planetary-mass companion is PSR
J1719$-$1438, an MSP with a period of 5.7~ms \citep{bbb+11a}. This
system is somewhat different to those described above in that it is
more akin to the binary systems (often known as ``black-widow'' systems)
which have very low-mass companions, e.g., PSR J0636+5129
\citep{slr+14}, but more extreme. In most black widow systems, the
radio emission is periodically eclipsed and they are believed to be
systems in which the companion is a stellar core being ablated by the
pulsar wind. They can have companion masses as low as 0.007 M$_\odot$
(about 7 Jupiter masses). PSR J1719$-$1438 does not show eclipses and
appears to be an ex-black-widow system in which the companion narrowly
survived the wind-blasting with a mass approximately equal to that of
Jupiter. \citet{bbb+11a} make a case for the companion having
a very high density, greater than 23~g~cm$^{-3}$, probably composed
mostly of carbon, leading to its moniker ``the diamond planet''.

Although all MSPs are believed to have passed through an evolutionary
phase where they had an accretion disk, it is rare for this accretion
disk to spawn a planetary system. The precise timing of MSPs limits
the number of planetary systems to those described above, only about
1\% of the population, although the relatively large intrinsic timing
noise of PSR B1937+21 can be intepreted as resulting from the
perturbations due to an asteroid belt surrounding the pulsar
\citep{scm+13}. 

\subsection{Tests of gravitational theories}\label{sec:grav}
The discovery of the first binary pulsar, PSR B1913+16, by
\citet{ht75a} marked a turning point in pulsar science. With its
short orbital period (7.75~h) and high orbital eccentricity (0.617),
it was immediately clear that precise timing of this pulsar would
allow detection of relativistic perturbations in the orbital
parameters. As mentioned above (\S\ref{sec:intro}), the short period
(59~ms) and large characteristic age ($10^8$~yr) of PSR B1913+16 indicated that
this pulsar was recycled. What turned out to be unusual was that the
Keplerian orbital parameters showed that the companion was massive
(minimum mass about 0.86 M$_\odot$) and probably another neutron
star. The two largest of the so-called ``post-Keplerian''
parameters\footnote{See \citet{sta03} for a description of
  the post-Keplerian parameterisation.}
periastron precession ($\dot\omega$) and relativitic time dilation,
usually described by the parameter $\gamma$, were detected at close to
their predicted values within a few years \citep{tfm79}. In Einstein's
general theory of relativity (GR), these two parameters depend on just
the masses of the two stars plus the Keplerian parameters (which were
well known). Consequently, the observation of these two parameters
allowed the masses to be derived. Both were close
to 1.4 M$_\odot$, confirming that PSR B1913+16 was a member of a
double-neutron-star system. Given these two masses, other relativistic
parameters could be predicted including, most importantly, orbital decay
due to emission of gravitational waves from the system. This too was
observed by \citet{tfm79}, fully consistent with the GR prediction.

\begin{figure}
\includegraphics[width=.95\columnwidth]{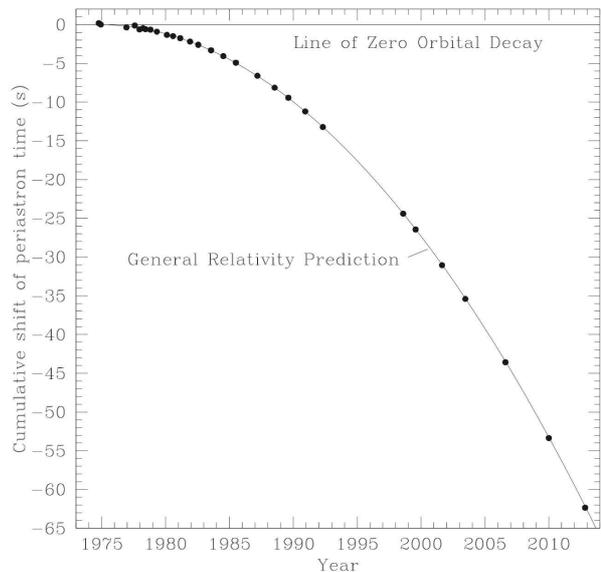}
\caption{Observed shift in the time of periastron passage relative to
  a constant orbital period for PSR B1913+16. The line is the
  predicted variation for orbit decay due to the emission of
  gravitational radiation from the system according to general
  relativity. \citep{wh16}  }\label{fg:B1913}
\end{figure}

Continued observations of this system have refined these parameters,
with the latest results (Figure~\ref{fg:B1913}) showing that the orbital decay
term is in agreement with the GR prediction (after compensation for
differential acceleration of the PSR B1913+16 system and the solar
system in the Galactic gravitational field) to better than
0.2\%. These observations also allowed detection of the relativistic
Shapiro delay parameters, $r$ and $s$, which depend on the orbit
inclination and companion mass, for the first time in this
system. The measured parameters are consistent with the GR
predictions although not very constraining since the orbital
inclination is close to $45^\circ$. 

The discovery of the Double Pulsar system, PSR J0737$-$3039A/B, at
Parkes \citep{bdp+03,lbk+04} made possible even more stringent tests of
GR. Its orbital period is only 2.4~h and the predicted periastron
advance, $16^\circ.9$~yr$^{-1}$, is more than four times that of PSR
B1913+16. Also, its orbital plane is almost exactly edge-on to us,
making the Shapiro delay large and easily measurable. Finally, the
companion star was observed as a pulsar (PSR J0737$-$3039B) with a long period (2.8~s) but
a much younger age than the A pulsar. Not only did this still unique
discovery allow a direct measurement of the mass ratio of the two
stars, it was fully consistent with the idea that the A pulsar was
(partially) recycled prior to the explosion of the companion star that
formed the B pulsar.

Continued timing observations using the Parkes, Green Bank and Jodrell
Bank telescopes have resulted in the measurement of five post-Keplerian
parameters, several to unprecedented levels of precision: $\dot\omega$ to
0.004\%, $\gamma$ to 0.6\%, the Shapiro delay terms $r$ and $s$ to 5\%
and 0.03\% respectively and the orbital period decay $\dot{P}_b$ to
1.4\% \citep{ksm+06}. Also, the mass ratio $R$ was measured to
0.1\%. The $s \equiv \sin i$ measurement implies an orbital inclination angle
of $88^\circ.7 \pm 0^\circ.7$. This is sufficiently close to edge-on
that the radiation from the A pulsar is eclipsed by the magnetosphere
of the B pulsar for just 30~s per orbit. Remarkably, high
time-resolution observations made with the Green Bank Telescope showed
that the eclipse is modulated at the rotation period of star B
\citep{mll+04}. Modelling of this eclipse pattern by Lyutikov \&
Thompson \citep{lt05}
allowed determination of the system geometry, including showing that
the rotation axis of B was inclined to the orbit normal by about
$60^\circ$. Even more remarkably, observations of the eclipse pattern
over a four-year data span gave a measurement of a sixth
post-Keplerian paramter, the rate of geodetic
precession, $(4^\circ.8\pm 0^\circ.7)$~yr$^{-1}$, consistent with the GR
prediction \citep{bkk+08}.

\begin{figure}
\includegraphics[width=.95\columnwidth,angle=270]{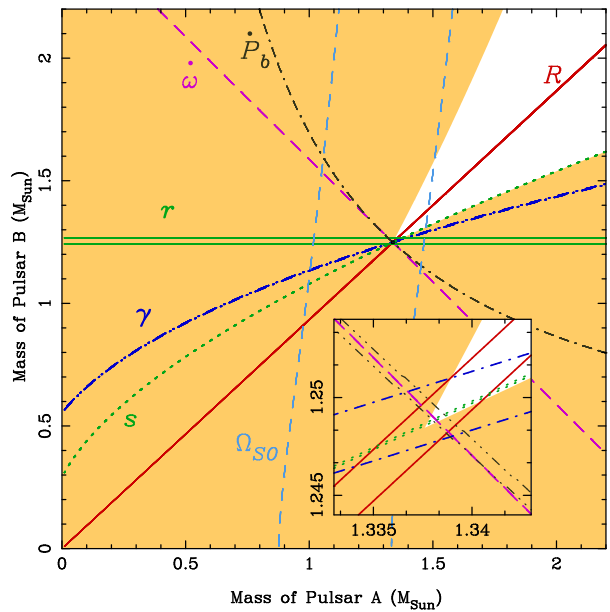}
\caption{Mass-mass diagram for the Double Pulsar system, updated to
  2015. Constraints from the various measurements in the context of
  general relativity are shown by pairs of lines representing the
  uncertainty in the measurement (not visible for the more precisely
  measured parameters). The yellow regions are permitted by the
  mass-function meaurement. The inset shows the region around the
  intersection of the various constraints. \citep{kra17}}\label{fg:J0737_m1m2}
\end{figure}

As shown in Figure~\ref{fg:J0737_m1m2}, all of these measurements can
be plotted on the so-called ``mass-mass'' diagram, a plot of companion
mass versus pulsar mass. Within the framework of GR, each
post-Keplerian measurement defines a zone on this plot within which
the two masses must lie. The Newtonian mass function and mass ratio
also define allowed regions. If GR is an accurate theory of gravity,
there will be a region on this diagram consistent with all
constraints, defining the masses of the two stars. Although it is
difficult to see, even on the inset, there is such a region on this
plot. These updated results verify that GR accurately descibes the
motion of the stars in the strong gravitational fields of this binary
system with a precision of better than 0.02\% \citep{kra17}.

Although GR has been incredibly successful as a theory of relativistic
gravity, passing every test so far with flying colours, it is by no
means the only possible theory of gravity. Departures from GR and the
equivalence principles that it is based on can be quantified in a
theory-independent way using the ``Parameterised post-Newtonian''
(PPN) parameters, see \citet{wil14}. Pulsars provide a variety of tests
that limit various combinations of these parameters, see
\citet{sta03}. Here, we describe just one such test: the effect of
``self-gravitation'' on the acceleration of objects in an external
gravitational field, a test of the Strong Equivalence Principle (SEP). This
test was first applied to solar-system dynamics by \citet{nor68b},
looking for a ``polarisation'' of the Moon's orbit in the direction of
the solar gravitational field. The test depends on the different
gravitational self-energy of the two binary components and so can be
tested using binary pulsars with low-mass companions and very low
eccentricity with the Galactic gravitational field as the polarising
agent. There is a large sample of such systems known and \citet{gsf+11}
analysed these to put a limit on the PPN parameter $\Delta$, which
measures the ratio of the gravitational and inertial masses,
effectively of the neutron star, of $4.3\times 10^{-3}$.

\begin{figure}
\includegraphics[width=.95\columnwidth]{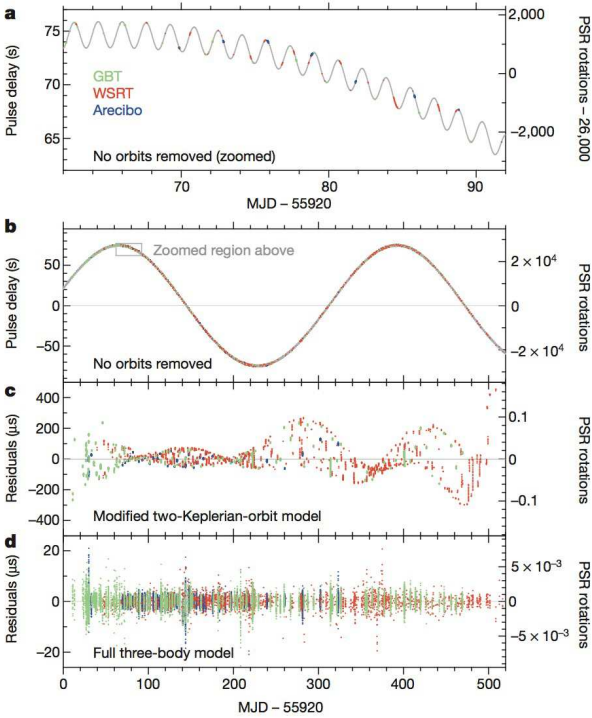}
\caption{Timing residuals for the PSR J0337+1715 system. Panel {\bf a} shows
  the timing signature of the inner binary orbit and is a zoomed-in
  portion of panel {\bf b} which is dominated by the timing delays due to
  the outer binary orbit. Panel {\bf c} shows the timing residuals from the
  fit of a standard two-orbit timing model to the more than 26,000
  observed pulse arrival times, and panel {\bf d} shows the residuals from a
  Markov-chain Monte Carlo fit of a full 3-body solution that includes
  the complex gravitational interactions between the three
  stars. \citep{rsa+14}}\label{fg:J0337}
\end{figure}

In 2014, \citet{rsa+14} announced the discovery of the fascinating
stellar triple system containing the pulsar J0337+1715, an MSP with
pulse period of 2.73~ms. The pulsar is in a relatively tight orbit,
orbital period 1.63~days, with a white dwarf of mass about
0.197~M$_\odot$. On a nearly co-planar orbit about this inner system, there
is a second white dwarf of mass about 0.41~M$_\odot$ and orbital
period about 327 days. The inner white dwarf has been optically
identified, leading to an estimate of the distance to the system,
about 1300~pc. Analysis of the complex interactions between the three
stars, illustrated in Figure~\ref{fg:J0337}, enabled determination of
the component masses (the neutron star has a mass close to
1.44~M$_\odot$) and the precise orbital inclinations for the two
systems which are equal to within 0$^\circ$.01 and close to
39$^\circ$.2.

An interesting aspect of this system is that it will provide a much
more sensitive test of the SEP than the
analysis of wide asymmetric binaries described above, through a potential
induced eccentricity of the inner orbit in the gravitational field of
the outer white dwarf \citep{rsa+14}. The gravitational field of the outer star at
the inner orbit is about six orders of magnitude greater than the
Galactic gravitational field in the local neighbourhood, leading to
the high expected sensitivity of the test. However, observations over several
orbits of the outer white dwarf will be required to separate any
SEP-violation effect from the intrinsic eccentricity of the inner
orbit, about $7\times 10^{-4}$. 

\section{The search for gravitational waves}\label{sec:pta}
Direct detection of the gravitational waves (GWs) predicted by
Einstein in 1916 \citep{ein16b} has been a major scientific goal over many
decades. With the remarkable discovery of a GW burst from two
coalescing 30-M$_\odot$ black holes by the LIGO-Virgo consortium in
September, 2015 \citep{aaa+16a}, this goal was
achieved. Laser-interferometer systems such as LIGO are sensitive to
GWs with frequencies in the range 10 -- 1000 Hz, the frequencies
expected from coalescing stellar-mass objects. The observed periods of
pulsars will also be perturbed by GWs passing through the Galaxy. But
because data spans of many years are required to reach the highest
precision, pulsar detectors are sensitive to much lower frequencies,
in the nanoHertz range. Likely astrophysical sources of such waves are
very different - most probably super-massive black-hole binary (SMBHB)
systems in the cores of distant galaxies. Studies of such waves are
therefore complementary to investigations using laser-interferometer
signals.

Pulsar GW-detection efforts depend on the great stability of MSP
periods. However, even the most stable pulsars can in principle have
intrinsic period irregularities, so observations of an ensemble of
MSPs, called a Pulsar Timing Array (PTA), is needed to detect GWs. The
detection method is based on searching for correlated signals among
the pulsars of a PTA which have the quadrupolar spatial signature
expected for GWs \citep{hd83}. Other correlated signals, such as those
produced by irregularities in the reference time standard, can also be
detected and distinguished from GWs by the different spatial response
pattern \citep{hcm+12}.

\begin{figure}
\includegraphics[width=.95\columnwidth]{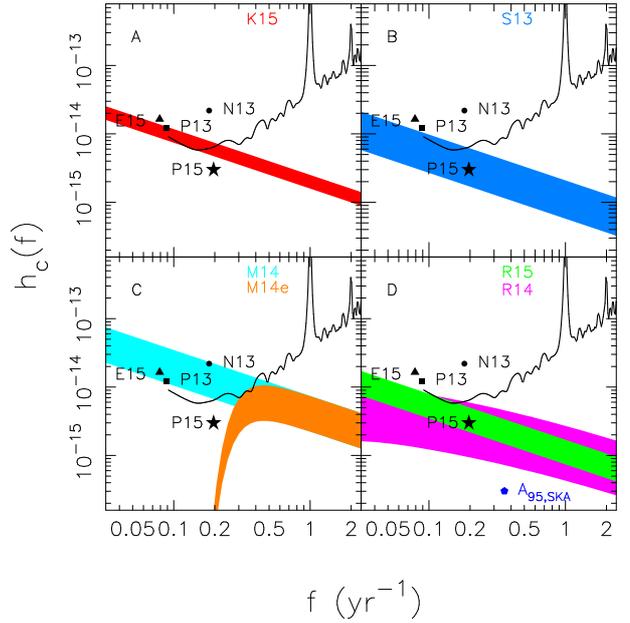}
\caption{Limits at 95\% confidence on a power-law gravitational-wave
  background for PTA observations. The four panels give predictions
  for the strength of such a background from various authors -- see
  the original paper for details. The current PPTA limit is marked by
  the star labelled P15. Limits from the NANOGrav collaboration
  \citep{dfg+13} (N13), the European Pulsar Timing Array \citep{ltm+15}
  and our earlier PPTA limit \citep{src+13} (P13) are marked. Also
  shown on each panel is the nominal frequency response of the PPTA to
  a monochromatic GW signal. In the lower right panel the anticipated
  sensitivity of a 5.5-year data set from the Square Kilometre Array
  \citep{jhm+15} is marked with a pentagon. \citep{srl+15}
}\label{fg:gwb_limit}
\end{figure}

Up to now, there has been no positive detection of nanoHertz GWs by a
PTA. However, limits on the strength of such a signal are beginning to
place interesting constraints on the source population and
properties. Although correlated signals among the pulsars of a PTA
must be observed to claim a detection, a limit on the strength of
nanoHertz GWs in the Galaxy can be obtained by placing limits on the
low-frequency signals in the modulation spectra of just the few best
pulsars in a PTA. The best such limit so far comes from analysis of data
from the Parkes Pulsar Timing Array (PPTA) \citep{mhb+13}.
\citet{srl+15} used 10~cm (3~GHz) observations of four of the most
stable PPTA pulsars to set a limit on the characteristic strain
amplitude $h_c < 3\times 10^{-15}$ at a GW frequency of 0.2 cycles per
year (6.3~nHz) of a power-law GW background (assumed spectral index
$-2/3$) in the Galaxy. This corresponds to an energy density of GWs at
this frequency that is a fraction $2.3\times 10^{-10}$ of the closure
energy density of the Universe. As Figure~\ref{fg:gwb_limit} shows,
the new limit rules out a number of models for SMBH evolution in
galaxies with high confidence.

In their paper, \citet{srl+15} identified enviromental effects
affecting the late evolution of SMBHBs as the most likely reason for
the current non-detection. If the SMBHB system loses energy to
surrounding stars or gas in its late evolution, it will pass through
this evolutionary phase more quickly than if GW emission were the sole
energy-loss process. This means that less energy will be emitted in
the form of GWs, thereby lowering the observed GW signal, particularly
at the lower observed frequencies (i.e., periods of decades)
\citep{rws+14}. However, the evidently low amplitude of nanoHertz GW in
the Galaxy can have other explanations. For example, the number density and/or
merger rate of SMBHs in the early Universe may be less than assumed,
see, e.g., \citet{cms+17}, or eccentricities of merging SMBHB may
be relatively large \citep{rws+14}. Both of these would have the effect of
reducing the GW amplitude at the low end of the PTA band where the
sensitivity is greatest.

Clearly, to achieve a detection of nanoHertz GWs and to begin to
explore their properties, increased PTA sensitivities are required.
As \citet{sejr13} pointed out, the most effective way to increase the
senstivity of a PTA is to increase the number of pulsars observed with
high timing precision. A start on this is being made by combining the
data sets of the three existing PTAs to form the International Pulsar
Timing Array (IPTA) \citep{vlh+16}. Observations with
future telescopes such as FAST \citep{nlj+11} and the SKA
\citep{jhm+15}, combined with existing data sets, 
will almost certainly result in a detection and open up an
era of nanoHertz-GW astronomy and astrophysics.

\section{Pulsar-based timescales}\label{sec:timescale}
As mentioned in the previous section, PTA data sets can also be used 
to investigate irregularities in the reference atomic timescales and
therefore to establish a pulsar-based timescale. International reference
timescales are currently based on a large number of atomic frequency
standards distributed across the world at many different time and
frequency standard laboratories, see \citet{app11}. These
measurements are collated at the Bureau International des Poids et
Mesures (BIPM)\footnote{www.bipm.org} in Paris to produce the
timescale TT(TAI) which is a continuous timescale with a unit that is
kept as close as possible to the SI second by reference to a few primary
caesium standards. 

\begin{figure}
\includegraphics[width=.95\columnwidth]{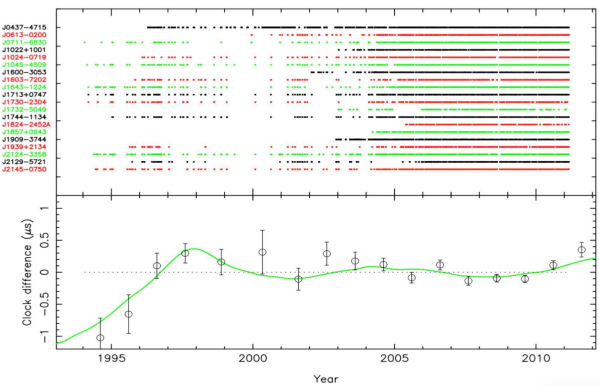}
\caption{The upper panel shows the data spans and sampling of the 20
  PPTA pulsars used to derive the common-mode signal shown in the
  lower panel. The timing analysis used TT(TAI) as a reference
  timescale and the line shows the quadratic-subtracted difference
  between TT(BIPM11) and TT(TAI). \citep{mghc17}
}\label{fg:psrtime}
\end{figure}

Although atomic frequency standards are improving all the time and
have reached incredible stabilities, of the order of a part in $10^{18}$
averaged over an hour or so for some optical lattice clocks, see
\citet{adh+15}, the long-term stability, over years and decades, of
these clocks is unknown. MSPs are highly stable clocks over intervals
of years and hence are well suited as an alternate
reference. Irregularities in the reference timescale would result in
inverse irregularities in the apparent period of all pulsars in a
PTA. This ``common-mode'' signal is relatively easy to identify and
separate from other perturbations in the pulsar periods
\citep{hcm+12}. Figure~\ref{fg:psrtime} shows the common-mode signal
derived from a reanalysis of the 20 PPTA pulsars used by \citet{hcm+12}
with TT(TAI) as a reference timescale. Because intrinsic pulsar
periods and slow-down rates are unknown and must be solved for as part
of the analysis, the pulsar timescale is insensitive to linear and
quadratic variations in the reference timescale. TT(TAI) is
``real-time'' and also contains known corrections to its rate. The
BIPM regularly reanalyses the atomic clock data to derive an improved
timescale TT(BIPMxx) where xx signifies the year of reanalysis -- this
is believed to be the most accurate long-term timescale available to us. In
Figure~\ref{fg:psrtime}, the line shows the quadratic-subtracted
difference between TT(BIPM11) and TT(TAI). Within the uncertainties,
the pulsar timescale accurately follows the known differences between
TT(BIPM11) and TT(TAI), demonstrating both that TT(BIPM11) is a more
uniform timescale than TT(TAI) and that pulsar timescales can have
comparable precision to the best international atomic timescales over
long time intervals. 

While current realisations of pulsar timescales are not quite at the
same level of stability as the best atomic timescales, they are
nevertheless valuable as an independent check on the long-term
stability of the atomic timescales. Firstly, they are completely
independent of terrestrial timescales and the terrestrial
environment. Secondly, they are based on entirely different physics,
rotation of a massive object, compared to the quantum-based atomic
timescales. Thirdly, the vast majority of MSPs will continue spinning
in a predictable way for billions of years, whereas the lifetime of an
atomic frequency standard is typically of the order of a decade. In
view of these points, the continued development and improvement of
pulsar timescales is certainly desirable and will happen as PTAs
improve. It is a nice thought that, in some sense, pulsar timescales
return time-keeping to its astronomical roots.

\section{Binary and stellar evolution}\label{sec:evol}
\begin{figure*}
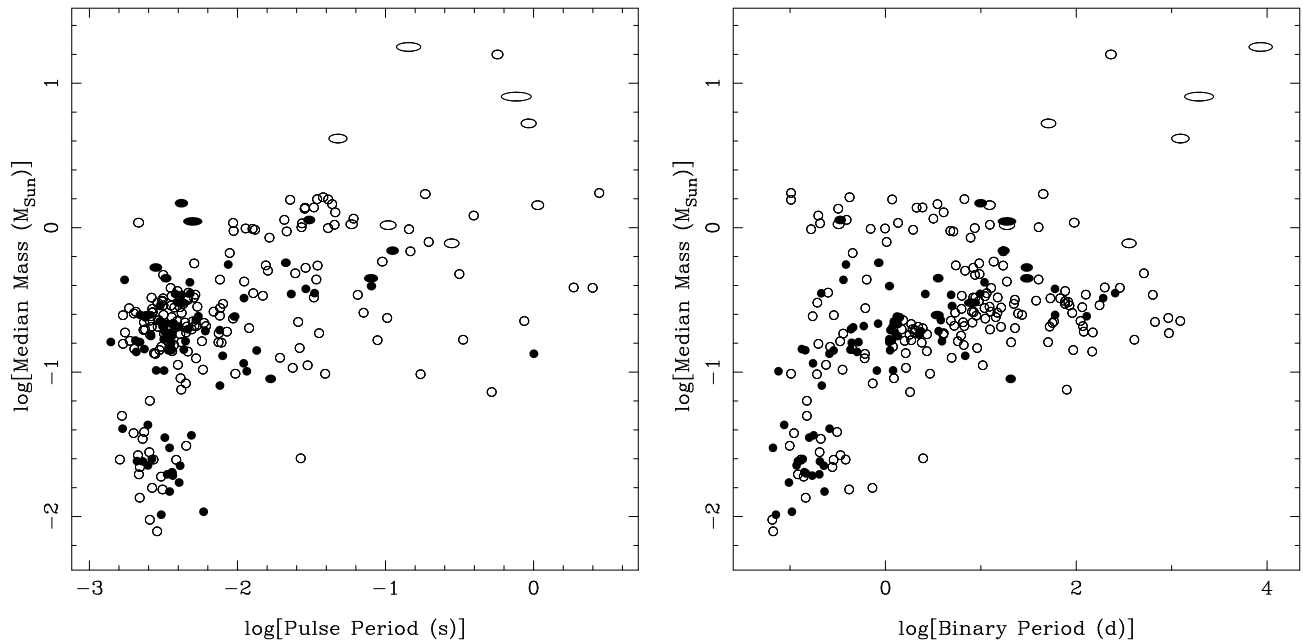

\begin{minipage}{170mm}
\begin{tabular}{cc}
\mbox{\includegraphics[width=84mm,angle=270]{p0_mc.ps}} &
\mbox{\includegraphics[width=84mm,angle=270]{pb_mc.ps}}
\end{tabular}
\caption{Median companion mass versus pulsar period (left) and orbital
  period (right) for pulsar binary systems. The orbit ellipticity is
  indicated by the ellipticity of the symbol and filled symbols
  indicate systems in globular clusters. Note that systems with
  planetary-mass companions are not included in the plots.}
\label{fg:mcomp}
\end{minipage}
\end{figure*}

As mentioned in the Introduction, the issue of how binary and
millisecond pulsars evolved to their present state has been a topic of
great interest right from the discovery of the first binary pulsar. In
the intervening 40 years or so the topic has become even more
fascinating with the discovery of MSPs in globular clusters, triple
systems, pulsars in orbit with very low-mass companions and eclipsing
systems. This diversity is illustrated in Figure~\ref{fg:mcomp} which
shows the median companion mass (computed from the binary mass
function with assumed orbit inclination $i=60^\circ$ and pulsar
mass $1.35$~M$_\odot$) of pulsar binary systems as functions of the pulsar
period $P$ and the orbital period $P_b$.

The most striking aspect of these plots is that systems with very
low-mass companions ($<0.08$~M$_\odot$) are concentrated at both short
pulsar periods (mostly $P<5$~ms) and short orbital periods
($P_b<1$~day). Most of these are black-widow systems, i.e., a
non-degenerate or partially degenerate stellar core that is being
irradiated and ablated by the pulsar wind. The ablated gas often results in
eclipses of the pulsar radio emission when the companion is close to
inferior conjunction, i.e., between the pulsar and us. PSR B1957+20,
discovered in 1988 \citep{fst88}, is the prototype for these systems. It
has a very low-mass companion (median mass 0.024~M$_\odot$) in a 9.17~h
circular orbit around an MSP with period 1.607~ms. The pulsar is
eclipsed for about 50~min each orbit with additional dispersive delays
before and after the eclipse as the pulsar enters and leaves the
gaseous wind from the companion. Initially, most of these PSR
B1957+20-like systems were
associated with globular clusters, but in recent years, largely thanks to the
great success of pulsar searches in the direction of unidentified {\it
  Fermi} $\gamma$-ray sources \citep{rap+12,cck+16}, the population is
roughly evenly divided between globular clusters and the Galactic
field (Figure~\ref{fg:mcomp}). 

Another population of binary systems with very short orbital periods
($P_b<1$~day) but somewhat larger companion masses, mostly between 0.1 and
0.3~M$_\odot$, can be identified on the $m_c$ -- $P_b$ plot. These are
the other type of ``spider'' pulsar known as redbacks
\citep{rob13}. The companion is a non-degenerate star that fills its
Roche lobe and is losing mass, partly as a result of heating by the
pulsar wind, resulting in extended and variable eclipses of the pulsar
radio signal -- see \citet{ccth13} for a discussion of the
formation processes of black-widow and redback eclipsing binaries. The
prototype redback (with hindsight) was PSR B1744$-$24A, the first pulsar
discovered in the globular cluster Terzan 5 \citep{lmd+90}, which has
the very short orbital period of 1.81~h, a median companion mass of
0.10~M$_\odot$ and is eclipsed for intervals that vary from orbit to
orbit from about 30\% of the orbital period to all of it. The two next
similar systems discovered (PSRs J0024$-$7204W \citep{clf+00} and
J1740$-$5340A \citep{dlm+01} also lie in globular clusters (47 Tucanae
and NGC 6397 respectively), suggesting that exchange interactions in
clusters is an important if not the only formation route for these
systems (see \citet{sp93} for a discussion of
stellar interactions in globular clusters).

However, the discovery of the  binary pulsar PSR J1023+0038 by
\citet{asr+09} changed this picture. This system has a 1.69-ms
pulsar in a 4.75-h circular orbit, with a companion of mass about
0.2~M$_\odot$, and shows deep and variable eclipses covering about
30\% of the period at frequencies about 1.4~GHz and more at lower
frequencies. These properties indicate a non-degenerate companion and
that the system is a member of the redback group. With Galactic
coordinates of $l=243^\circ.5$, $b=45^\circ.8$ and a distance of about
1.4~kpc, the PSR J1023+0038 system lies far from any globular cluster
and, hence, an origin in a cluster is very unlikely. This showed that
such systems could result from evolution of a binary system without the
need to invoke exchange interactions; possible evolutionary paths are
discussed by \citet{ccth13}. Since then, the {\it Fermi}-related
searches \citep{rap+12,cck+16} have uncovered many more redback systems
in the Galactic field, such that they now out-number the
globular-cluster redbacks (see right panel of Figure~\ref{fg:mcomp}),
reinforcing this conclusion.

Compared to PSR B1913+16, the next two binary pulsars discovered, PSR
B0820+02 \citep{mncl80} and PSR 0655+64 \citep{dbtb82} had very
different properties. PSR B0820+08 has a relatively long pulsar
period, 0.865~s, a very long orbital period, about 3.3 years, a median
companion mass of 0.22~M$_\odot$ and no sign of eclipses. PSR B0655+64
has a pulse period of 0.196~s and is in a circular orbit with a period
of just over one day, giving a median companion mass of
0.79~M$_\odot$. Again, there is no sign of any eclipse. These
properties suggest that the companions in these two systems are
compact degenerate stars and this was later confirmed by optical
identifications of white dwarf companions \citep{kr00,kul86}. These
were the first of many discoveries of pulsars with white dwarf
companions, some believed (or known) to be helium white dwarfs with
masses in the range 0.1 -- 0.3~M$_\odot$, and others with more massive
carbon-oxygen (CO) or oxygen-neon-magnesium (ONeMg) white dwarf
companions. Between them these white-dwarf systems account for about
60\% of the known binary pulsars, with about three-quarters of them
having He white dwarf companions. As Figure~\ref{fg:mcomp}
illustrates, many of them have highly recycled pulsars (pulse period
of a few milliseconds or less) and most have relatively long orbital
periods. The lower-mass He white dwarf systems are believed to have
evolved from low-mass X-ray binary (LMXB) systems whereas the more
massive systems with a CO white dwarf are probably formed in
intermediate-mass X-ray binary (IMXB) systems - see
\citet{tlk12} for a detailed discussion of these evolutionary
processes.

Another important group of pulsars is those having neutron star
companions. These double-neutron-star systems have intermediate pulsar
periods, most between 20~ms and 100~ms, indicating a short accretion
and spin-up phase. This is consistent with the faster evolution of the
more massive stars needed to form neutron stars by collapse of the
stellar core at the end-point of their evolution. Their orbital
periods range between 0.1 day (for the Double Pulsar) and several tens
of days and they have relatively high eccentricities
(Figure~\ref{fg:mcomp}). Only one, B2127+11C, lies in a globular
cluster (M15).

Finally, we have the binary systems with massive main-sequence
companions. The prototype is PSR B1259$-$63, a 47.7~ms pulsar in a 3.4~year
highly eccentric orbit ($e \sim 0.808$) around a 20~M$_\odot$ Be star
LS 2883 \citep{sjm14}. The pulsar is eclipsed in the radio for about
100 days around periastron and emits high-energy (X-ray and $\gamma$-ray) unpulsed
emission \citep{cna+09,aaa+11} as it passes through the circumstellar
disk of the Be star. Only a handful of similar systems are known and
they are shown in the upper-right region of both panels in
Figure~\ref{fg:mcomp}.

The idea that binary (and single) MSPs get their short periods and low
magnetic fields by a recycling process occuring in X-ray binary
systems, was first discussed by \citet{sb76} and
\citet{sv82} in the context
of PSR B1913+16, and then extended to account for the properties of
the first MSP, PSR B1937+21, by 
\citet{rs82} and \citet{acrs82}. Since
that time, the field has become incredibly richer with the discovery
of the diverse types of binary and triple systems described in the
preceding sections. This diversity can be accommodated within the
recycling model by invoking different initial conditions (component
masses, orbital periods etc.) and different environments, e.g.,
globular clusters or Galactic field, for the progenitor binary systems
- see \citet{bv91} and \citet{tv06} for extensive reviews of the evolution of compact
X-ray binary systems.

Many observations have supported the recycling hypothesis. For
example, the properties of the second-formed B pulsar in the Double
Pulsar system are completely in accord with expectations from the
recycling model \citep{lbk+04}. But much more direct evidence has
recently been found with the discovery of the ``transitional''
systems, PSR J1023+0038 \citep{asr+09}, PSR J1824$-$2452I (M28I)
\citep{pfb+13} and PSR J1227$-$4853 \citep{rrb+15}. In the first case, a
relatively bright MSP with an orbital period of 0.198 days and wide
eclipses, was discovered in a wide-area 350-MHz survey at Green
Bank. Within the uncertainties, the MSP is coincident with a
previously known solar-type star that showed brightness variations
with a 0.198-day periodicity \citep{ta05}, thereby clinching the
identification. This combination of properties identified the system
as a redback, with the solar-type star being the accretion donor in
the earlier X-ray accretion phase. What makes this system especially
interesting (besides not being in a globular cluster) is that optical
spectra taken in 2000 -- 2001 showed a very different blue and
double-lined spectrum characteristic of an accretion disk
\citep{sfs+03}. This suggests that the system is bistable, oscillating
between an accretion IMXB phase and a MSP redback phase. This was
confirmed by observations of a sudden quenching of the pulsar radio
emission and a coincident return to the optical and X-ray behaviour
expected for an IMXB in 2013 \citep{pah+14}. Also, a coincident sudden
increase in the $\gamma$-ray flux from the source was reported by
\citet{sah+14}.

In 2013, \citet{pfb+13} also reported clear evidence for the
connection between recycled MSPs and X-ray binary systems. In April,
2013, {\it XMM Newton} observations revealed coherent pulsations with
a period of 3.931~ms in an X-ray transient source detected a few weeks
earlier by the X-ray satellite {\it INTEGRAL}, marking this as an
accretion-powered MSP similar to SAX J1808.4$-$3658 \citep{hpc+09}. It
was quickly realised that this is the same source as the binary radio
pulsar J1824$-$2452I in M28. Archival observations from 2008 showed
that the radio pulsar was detected just two months before a {\it
  Chandra} X-ray detection, showing that the change of state was
rapid. This was confirmed by X-ray and radio observations in mid-2013
showing that the system swapped between X-ray and radio-pulsar states
on timescales as short as a few days. 

The PSR J1227$-$4853 system appears to be similar to that of PSR
J1023+0038. Both are members of systems in which there was a rapid
transition between an accreting binary X-ray
source and a radio MSP. The binary X-ray source, XSS J12270$-$4859,
was observed to transition to a quiet state with none of the optical
or X-ray signatures of an accretion disk in late 2012
\citep{bph+14}. A radio search at the X-ray position using the Giant
Metrewave Radio Telescope (GMRT) in India and subsequent timing
observations using the GMRT and the Parkes 64-m radio telescope
revealed a binary eclipsing 1.686~ms pulsar with the same orbital
period as the X-ray source, 6.91~days \citep{rrb+15}. The binary
parameters are consistent with a companion mass in the range 0.17 --
0.46~M$\odot$, suggesting that the system is a member of the redback
group.

These observations of binary systems which transition between X-ray
accretion-powered systems to rotation-powered radio MSPs
provide the best evidence yet that MSPs are indeed old and slowly rotating
neutron stars that have been spun up, or recycled, by accretion in an X-ray
binary system. 

Observations of binary MSPs, including those detected in X-ray
accreting and burst systems, also provide important constraints on the
structure and the equation of state for neutron stars and other hybrid
or quark stars (for recent reviews see \citet{hz16} and
\citet{of16}). The most important constraints come from the large
pulsar masses implied by either Shapiro delay measurements, as for PSR
J1614$-$2230 \citep{dpr+10,fpe+16} and PSR J1946+3417 \citep{bfk+17}, or
by optical identifcation and mass measurement of the companion star as
for PSR J0348+0432 \citep{afw+13} and PSR J1012+0507
\citep{ato+16}. These four systems have estimated pulsar masses of
$1.928\pm 0.017$~M$_\odot$, $1.828\pm 0.022$~M$_\odot$, $2.01\pm
0.04$~M$_\odot$ and $1.83\pm 0.11$~M$_\odot$, respectively, all above
1.8~M$_\odot$. These high masses are inconsistent with equations of
state that are ``soft'' at high densities \citep{dpr+10,hz16}. On the
other hand, many ``hard'' equations of state are ruled out with X-ray
measurements that constrain the stellar radius to values around 10 --
12 km \citep{of16}.

\begin{figure}
\includegraphics[width=.95\columnwidth]{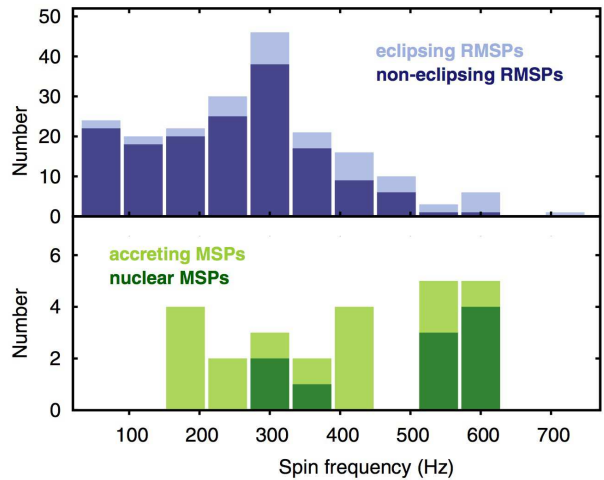}
\caption{Observed distribution of spin frequency for the fastest MSPs
  including radio eclipsing and non-eclipsing MSPs, X-ray accretion-powered MSPs and
  X-ray nuclear-powered or burst MSPs. \citep{ptrt14}
}\label{fg:spin}
\end{figure}

Another important aspect of MSP properties that impacts on
neutron-star physics is the maximum observed spin
rate. Figure~\ref{fg:spin} shows the observed spin frequency
distribution of the fastest MSPs according to source type. Despite the
discovery of several eclipsing MSPs with pulse
frequencies above 500 Hz in radio searches of {\it
  Fermi} sources in the past few years, the spin rate record (716~Hz)
is still held by PSR J1748$-$2446ad in the globular cluster Terzan 5,
found more than a decade ago \citep{hrs+06}. Searches for short-period
(sub-millisecond) MSPs at both radio or X-ray frequencies have not
been instrumentally limited since that time or even before. However, there
is likely to be a selection against short-period pulsars in
lower-frequency radio searches because of absorption and scattering in the
circumstellar plasma of eclipsing binary radio pulsars -- PSR
J1748$-$2446ad was found in a 2-GHz search using the Green Bank
Telescope. It is clear that there is a physical mechanism limiting the
maximum spin frequency of neutron stars in accreting binary systems to
something around 700~Hz despite most neutron-star equations of state
allowing maximum spin frequencies of more than 1~kHz
\citep{lp07}. Gravitational ``r-mode'' instabilities were suggested by
\citet{bil98} as a mechanism, but these appear to be too
efficient and must be suppressed in some way to allow the higher
observed spin frequencies \citep{cgk17}.  It is possible that spin
rates are naturally limited by a balance between spin-up and spin-down
torques \citep{phd12,bc17}.

\section{Conclusion}\label{sec:concl}
The discovery of the first binary pulsar in 1975 and the first
MSP in 1982 opened the door to a huge diversity of
research fields, ranging from tests of theories of relativistic
gravitation to understanding the evolutionary pathways that could
lead to such objects. In this brief review, I have just skimmed the
surface of a few of these research fields, omitting mention of many
altogether. I hope though that this has been enough to illustrate the
enormous power and potential of the study of millisecond pulsars and
pulsar binary systems. In this volume celebrating the contributions of
Srinivasan to astrophysics, I have highlighted his contributions to
the topics of MSP and binary evolution. With colleagues including Bhattacharya,
Radhakrishnan and van den Heuvel, he has made ground-breaking
contributions to the field right from the early days following the
discovery of the first binary pulsar. Many of these contributions have
proved prescient and are part of the present-day understanding of these
topics. We thank him for his many insightful contributions over the
past 40 years.

%\balance
 
%%Use section* for acknowledgements
\section*{Acknowledgements}
The NASA Astrophysics Data System
(http://www.adsabs.harvard.edu/) and the ATNF Pulsar Catalogue (V1.56,
www.atnf.csiro.au/research/pulsar/psrcat, \citep{mhth05}) were used extensively in the
preparation of this review.

%%use \balance somewhere in the left column of the last page to balance the two columns in the end page

%\bibliographystyle{mnras}
%\bibliography{journals,modrefs,psrrefs,crossrefs}

%%References section
%\begin{thebibliography}{99} 
%\bibitem{latex1} 
%Clark, D. H., Caswell, J. L. 1976, {\em MNRAS}, {\bf 174}, 267. 
%\bibitem{latex2} 
%Dickey, J. M., Salpeter, E. E., Terzian, Y. 1978, {\em Astrophys. J. Suppl. Ser.}, {\bf 36}, 77.
%\bibitem{latex3} 
%Radhakrishnan, V. 1980, in {\em Non-Solar Gamma Rays (COSPAR)}, Eds R. Cowsik and R. D. Wills, Pergamon Press, Oxford, p. 163.
%\bibitem{latex4} 
%Zwicky, F. 1957, {\em Morphological Astronomy}, Springer-Verlag, Berlin, p. 258.
%\end{thebibliography}

\end{document}